\documentclass[aps,prd,notitlepage,showpacs,nofootinbib,superscriptaddress]{revtex4-1}

\usepackage{graphicx}
\usepackage[utf8]{inputenc} 

\usepackage{amsmath}
\usepackage{amssymb}
\usepackage{braket}
\usepackage{float}
\usepackage{comment}
\usepackage{slashed}
\usepackage[normalem]{ulem}
\usepackage{adjustbox}
\usepackage{multirow}

\usepackage{caption}
\usepackage{subcaption}
\usepackage{amsmath}
\usepackage{amssymb}

\usepackage[usenames,dvipsnames]{color}
\usepackage[colorinlistoftodos]{todonotes}
\usepackage[colorlinks=true,citecolor=darkred,urlcolor=darkred, pdfborder={0 0 0}]{hyperref}
\usepackage[normalem]{ulem}

\definecolor{darkred}{rgb}{0.6,0,0}
\definecolor{darkbrown}{rgb}{0.39,0.26,0.12}
\usepackage[colorinlistoftodos]{todonotes}

\definecolor{linkcolor}{rgb}{0,0,0.5}
\newcommand {\ignore}[1]{}

\def\be{\begin{equation}}
\def\ee{\end{equation}}

\def\gsim{\raise0.3ex\hbox{$\;>$\kern-0.75em\raise-1.1ex\hbox{$\sim\;$}}}
\def\lsim{\raise0.3ex\hbox{$\;<$\kern-0.75em\raise-1.1ex\hbox{$\sim\;$}}}

\usepackage{soul}

\definecolor{mightnightblue}{RGB}{25,25,112}

\definecolor{brown}{rgb}{0.59, 0.29, 0.0}

\def\21{$\mathrm{SU(2)_L \otimes U(1)_Y}$}

\usepackage{booktabs}
\usepackage{tabularx}
\usepackage{multirow}
\usepackage{makecell}
\usepackage{siunitx}

\begin{document}
\bibliographystyle{unsrt}   % needed for refs and hyperlinks %% utphys.bst style file is also needed with this %%%%%%%%%%%%

\title{
\color{BrickRed} Probing Dirac Heavy Neutral Leptons through EFT-Induced Meson Decays} 

\author{Sai Gade}\email{saishagade@gmail.com}
\affiliation{School of Physics, University of Hyderabad, Hyderabad - 500046, India}

\author{Sanjoy Mandal}\email{smandal@kias.re.kr}
\affiliation{Korea Institute for Advanced Study, Seoul 02455, Korea}

\author{Manimala Mitra}\email{manimala@iopb.res.in}
\affiliation{Institute of Physics, Sachivalaya Marg, Bhubaneswar 751005, India}
\affiliation{Homi Bhabha National Institute, BARC Training School Complex,
Anushakti Nagar, Mumbai 400094, India}

\author{Sushant K. Raut}\email{sushant.raut@krea.edu.in}
\affiliation{Krea University, Sri City, India 517646}
%%%%%%%%%%%%%%%%%%%%%%%%%%%%%%%%%%%
\begin{abstract}
We investigate dimension-six operators involving pairs of Dirac-type heavy neutral leptons (HNLs) at beam-dump and far-forward experiments. In the \(\nu\)SMEFT framework, lepton-number-conserving four-fermion operators couple HNL pairs to quarks and induce two and three-body charm and bottom meson decays. We study the \(c\to u\), \(b\to d\), and \(b\to s\) transitions at FASER2, FACET, SHiP, and two FPF@FCC configurations. HNL production depends on the cut-off scale $\Lambda$, while the lifetime depends on the active–sterile mixing \(|V_{eN}|^2\). Sizable production can therefore coexist with the small mixing needed for HNLs to reach distant detectors. We project sensitivities in the \((M_N,|V_{eN}|^2)\) and \((M_N,\Lambda)\) planes, accounting for two and three-body decays and constraints from invisible and semivisible meson decays. SHiP is particularly sensitive to \(c\to u\), while the larger FCC-LLP2 configuration probes \(b\to d,s\). We find that EFT-induced pair production extends long-lived HNL searches to very small mixing and complements other probes of the \(\nu\)SMEFT heavy-neutrino sector.
\end{abstract}

\maketitle

%\tableofcontents

%%%%%%%%%%%%%%%%%%%%
\section{Introduction}
%%%%%%%%%%%%%%%%%%%%%%%%%%%
Heavy neutral leptons (HNLs) constitute a well-motivated extension of the Standard Model (SM) and arise naturally in mechanisms that account for the observed nonzero neutrino masses~\cite{Schechter:1980gr,Minkowski:1977sc,Gell-Mann:1979ijt,Mohapatra:1979ia,Mohapatra:1986aw,Mohapatra:1986bd,Yanagida:1979as}. Depending on the underlying neutrino-mass mechanism and its lepton-number structure, the heavy states may be Majorana, Dirac, or pseudo-Dirac fermions. In particular, approximate lepton-number conservation can lead to nearly degenerate heavy states whose phenomenology approaches that of a Dirac HNL~\cite{Wolfenstein:1981kw,Petcov:1982ya,Valle:1982yw,Batra:2023mds}. Searches for these particles therefore offer an important opportunity to investigate the origin of neutrino masses and the interactions of new neutral fermions.

In conventional HNL searches, both production and decay are typically governed by active--sterile neutrino mixing, $V_{\alpha N}$, with $\alpha=e,\mu,\tau$~\cite{Das:2024kyk,Das:2023tna,Das:2018usr,Cottin:2018kmq,Cottin:2018nms,ATLAS:2019kpx,CMS:2022fut,Mitra:2016kov,Banerjee:2015gca,Das:2017pvt,Deppisch:2018eth,Antusch:2019eiz,Antusch:2018bgr,Antusch:2017hhu,Antusch:2016ejd,Antusch:2016vyf,Antusch:2015mia,Drewes:2024bla,Abdullahi:2022jlv,Drewes:2019byd,Drewes:2013gca,Chakraborty:2018khw}. Small mixing suppresses the HNL decay width and allows it to travel macroscopic distances before decaying, giving rise to displaced-vertex signatures. However, the same mixing suppression also reduces the production rate. This correlation limits the reach of conventional long-lived-particle (LLP) searches, particularly at very small mixing~\cite{Kling:2018wct,Drewes:2018gkc,SHiP:2018xqw,Helo:2018qej,Chun:2019nwi,Mandal:2017tab,Mandal:2016hpr,Alimena:2019zri,Drewes:2018xma,Curtin:2018mvb,Drewes:2018gkc,Alekhin:2015byh,Wang:2025esc}. 

Effective field theory (EFT) provides a systematic way to describe additional HNL interactions that need not be suppressed by active--sterile mixing. In the HNL extended Standard Model Effective Field Theory ($\nu$SMEFT), higher-dimensional operators involving gauge-singlet fermions can open new production channels and connect HNL searches to a new-physics scale~\cite{Biswas:2024gtr,Chun:2024mus,Mitra:2022nri,Beltran:2024twr,Beltran:2023nli,Cottin:2022nwp,Beltran:2022ast,Beltran:2021hpq,Cottin:2021lzz,Hirsch:2020klk}. We focus on lepton-number-conserving dimension-six four-fermion operators coupling a pair of Dirac HNLs to quarks. These interactions induce HNL pair production through two-body ($P\to N\bar N$) and three-body ($P\to XN\bar N$) meson decays, where $P$ is the parent pseudoscalar meson and $X$ is a lighter pseudoscalar or vector meson. Both two and three-body channels can contribute appreciably, with their relative importance depending on the HNL mass and available phase space.

A distinctive feature of this scenario is the separation of HNL production and decay~\cite{DeVries:2020jbs,Cottin:2021lzz,Beltran:2021hpq,Beltran:2022ast,Beltran:2025ilg,A:2026srg,KA:2026czk,A:2025ygb}. The EFT-induced meson decay rates depend on the effective coupling $c_{\mathcal O}/\Lambda^2$, where $\Lambda$ is the cutoff scale, whereas the subsequent decay of the lightest HNL into SM particles is controlled by the active--sterile mixing $|V_{\alpha N}|^2$. The pair-HNL operators considered here do not, by themselves, mediate the decay of the lightest HNL into SM states. Consequently, sizable HNL production can coexist with sufficiently small mixing to yield macroscopic decay lengths. This decoupling of production and decay makes the framework especially attractive for searches at distant LLP detectors and opens parameter regions that are challenging for mixing-driven production mechanisms.

High-intensity proton beam-dump experiments and far-forward detectors at hadron colliders provide complementary settings for these searches. Copious production of charm and bottom hadrons supplies potential HNL sources, while downstream decay volumes can detect visible decays of particles that traverse substantial shielding. The resulting sensitivity depends on the parent-meson spectra, the HNL boost and lifetime, the distance from the production point, and the detector's geometrical acceptance. These factors vary significantly between fixed-target facilities and forward detectors at high-energy colliders.

In this work, we investigate five such proposed detector configurations. At the High-Luminosity Large Hadron Collider (HL-LHC), we consider FASER2~\cite{Feng:2017uoz,Feng:2022inv,Anchordoqui:2021ghd} and FACET~\cite{Cerci:2021nlb}, which target LLPs produced in the forward direction of proton--proton collisions. We also study the SHiP experiment~\cite{Alekhin:2015byh,SHiP:2020vbd,SHiP:2021nfo} at the CERN Super Proton Synchrotron, where a high-intensity 400~GeV proton beam incident on a fixed target generates abundant heavy-flavor hadrons. To explore the opportunities at higher collision energies, we further consider the Forward Physics Facility at the Future Circular Collider (FPF@FCC)~\cite{FCC:2018byv,FCC:2018vvp,Aleksa:2019pvl,MammenAbraham:2024gun}, associated with $\sqrt{s}=100$~TeV proton--proton collisions. Following Ref.~\cite{MammenAbraham:2024gun}, two benchmark configurations are included: FCC-LLP1 (FPF1), with dimensions $5\times5\times50~\mathrm{m}^3$, and FCC-LLP2 (FPF2), with dimensions $20\times20\times400~\mathrm{m}^3$. Their different production rates, baselines, and fiducial volumes allow us to assess the complementary reach of present-generation and future experimental concepts.

We calculate the contributions of two and three-body $D$ and $B$-meson decays to Dirac HNL pair production, accounting for meson spectra, propagation and decay probabilities, and detector geometry. For definiteness, we assume dominant mixing with the electron neutrino and treat $M_N$, $|V_{eN}|^2$, and $c_{\mathcal O}/\Lambda^2$ as independent parameters. We present projected sensitivities in the $(M_N,|V_{eN}|^2)$ and $(M_N,\Lambda)$ planes and compare them with existing bounds. In this way, our analysis examines how EFT-induced pair production can extend the reach of LLP searches beyond the conventional assumption that production and decay are both governed by active--sterile mixing.

The remainder of this paper is organized as follows.
In Sec.~\ref{sec:model}, we introduce the EFT
framework for Dirac HNLs.
Sec.~\ref{sec:meson} discusses the relevant dimension-six operators
and the meson decay channels through which HNLs can be produced.
In Sec.~\ref{sec:HNL_decay}, we examine the decay modes of Dirac HNLs.
Sec.~\ref{sec:simulation} describes the experimental setups and
numerical simulation methodology adopted in our analysis.
We present and discuss our numerical results in
Sec.~\ref{sec:results}, before summarizing our main findings
and concluding in Sec.~\ref{sec:conclusion}.
%%%%%%%%%%%%%%%%%%%%%%%%%%%
%%%%%%%%%%%%%%%%%%%%%%%%%%
\section{Heavy Dirac neutrino EFT}
\label{sec:model}
%%%%%%%%%%%%%%%%%%%%%%%%%%%%
HNLs are commonly introduced to explain the observed nonzero masses of active neutrinos. In the conventional type-I seesaw mechanism~\cite{Schechter:1980gr,Minkowski:1977sc,Gell-Mann:1979ijt,Mohapatra:1979ia,Mohapatra:1986aw,Mohapatra:1986bd,Yanagida:1979as}, the HNLs are Majorana particles. However, in low-scale neutrino-mass models such as the inverse~\cite{Mohapatra:1986bd} and linear seesaw~\cite{Akhmedov:1995ip,Akhmedov:1995vm} mechanisms, the heavy states naturally form quasi-Dirac pairs~\cite{Wolfenstein:1981kw,Petcov:1982ya,Valle:1982yw}. In the exact lepton-number-conserving limit, the mass splitting between the two members of each pair vanishes, and they combine to form a Dirac HNL.
%%%%%%%%%%%%%%%%%%%%%%%%%%%%
\par The $\nu$SMEFT extends the Standard Model effective field theory by including the gauge-singlet field $N$. It comprises a complete set of higher-dimensional operators constructed from the SM fields and $N$ that are invariant under the SM gauge group. The corresponding effective Lagrangian can be expressed as~\cite{Aparici:2009fh,Bhattacharya:2015vja,Liao:2016qyd,Li:2021tsq,Bolton:2025tqw,Cottin:2021lzz,Mitra:2022nri,Beltran:2022ast},
\begin{align}
\label{eq:vSMEFT}
\mathcal{L}
={}&
\mathcal{L}_{\mathrm{SM}}
+\overline{N}i\slashed{\partial}N
-\left[
\overline{L}Y_\nu \widetilde{H} N
+\frac{1}{2}\overline{N^c}M N
+\mathrm{h.c.}
\right]
+\frac{1}{\Lambda^{d-4}}\sum_i C_i^{(d)}O_i^{(d)},
\end{align}
where $H$ is the SM Higgs doublet and  $\tilde{H}=i\sigma_2 H^{*}$, $L$ is the SM lepton doublet and $Y_\nu$ denotes the neutrino Yukawa coupling, $M$ is the Majorana mass matrix of the right-handed neutrinos, and $C_i^{(d)}$ are the Wilson coefficients associated with the dimension-$d$ operators $O_i^{(d)}$. The cutoff scale of the EFT is denoted by $\Lambda$.  Eq.~\ref{eq:vSMEFT} can be considered as the type-I seesaw version of $\nu$SMEFT. Following electroweak symmetry breaking~(EWSB), the Yukawa interaction $\overline{L}Y_\nu\widetilde{H}N$ and the Majorana mass term $\frac{1}{2}\overline{N^c}M N$ generate the active-neutrino mass matrix through the type-I seesaw mechanism. The neutrino masses may receive additional contributions from the higher-dimensional operators appearing in Eq.~\ref{eq:vSMEFT}. This will introduce the following mixing between active and heavy neutrino,
%%%%%%%%%%%%%%%%%%%%%%%%%%%%%%%%%%%%
\begin{align}
\label{eq:mixing_Majorana}
\begin{pmatrix}
\nu \\
N^c
\end{pmatrix} = 
P_L\begin{pmatrix}
1 - \frac{1}{2}\Theta\Theta^\dagger & \Theta \\
- \Theta^\dagger & 1 - \frac{1}{2}\Theta^\dagger\Theta
\end{pmatrix}\begin{pmatrix}
\nu' \\
N'
\end{pmatrix} \,; \quad \Theta_{\alpha i} \equiv V_{\alpha N_i} = \frac{v(Y_\nu)_{\alpha i}}{\sqrt{2}M_i} \,,
\end{align}
%
%%%%%%%%%%%%%%%%%%%%%%%%%%%%%%%%%
The resulting massive states are Majorana fermions ($N' = N^{\prime c}$) and the active-sterile mixing $V_{\alpha N_i}$ couples $N_i'$ to the SM via the charged and neutral currents. The phenomenology of Majorana HNLs with non-zero active-sterile
mixing has been reviewed in Refs.~\cite{Atre:2009rg, Bolton:2019pcu, Fernandez-Martinez:2023phj,Chun:2024mus,Mitra:2022nri,Das:2018usr}.
%%%%%%%%%%%%%%%%%%%%%%%%%%%
\par Here we consider instead the lepton number conserving~($\Delta L=0$) limit of
Eq.~\ref{eq:vSMEFT}, focusing only on the operators relevant to the current
study. A heavy Dirac neutrino cannot be constructed from a single right-handed Weyl field $N$. One must introduce an additional right-handed singlet $S$, with both fields carrying the same conserved lepton number, $L[N]=L[S]=1$. It allows the lepton-number-conserving Dirac mass term $M\bar{S} N$, while Majorana mass terms for $N$ and $S$ are forbidden. For simplicity, we assume an equal number of $N$ and $S$ fields~\footnote{Although the number of RHNs can be arbitrary, we adopt a symmetric setup for concreteness, introducing one $N$ and one $S$ for each SM fermion generation. For simplicity, however, the numerical analysis is performed using a single generation of HNLs.}. An exact $U(1)_L$, $U(1)_{B-L}$, or an appropriate discrete symmetry forbids Majorana masses and all $\Delta L=2$ operators. In the $\Delta L = 0$ limit, the effective Lagrangian becomes
\begin{align}
\label{eq:vSMEFT_S}
\mathcal{L} = \mathcal{L}_{\text{SM}} + \bar{N}i\slashed{\partial}N + \bar{S}i\slashed{\partial}S - \bigg[\bar{L}Y_\nu N\tilde{H} + \bar{S} M N + \text{h.c.} \bigg] + \frac{1}{\Lambda^{d-4}}\sum_{d>6,\, i} C_i^{(d)} Q_i^{(d)}\,,
\end{align}
%%%%%%%%%%%%%%%%%%%%%%%%%%%%%%%
where the sum is over $\Delta L = 0$ operators constructed from SM degrees of freedom plus $N$ and $S$. As we are considering Dirac HNL, we focus only on $B-L$ invariant dimension six operators~\footnote{Note that the dimension five operators involving Dirac HNLs are absent due to the $B-L$ symmetry}.  Active-sterile mixing is again induced in the broken phase, up to $\mathcal{O}(\Theta^2)$, as 
\begin{align}
\label{eq:mixing_Dirac}
\begin{pmatrix}
\nu \\
N^c \\
S
\end{pmatrix} = 
P_L\begin{pmatrix}
1 - \frac{1}{2}\Theta\Theta^\dagger & 0 & \Theta \\
0 & 1 & 0 \\
- \Theta^\dagger & 0 & 1 - \frac{1}{2}\Theta^\dagger\Theta
\end{pmatrix}\begin{pmatrix}
\nu' \\
N^{\prime c} \\
N'
\end{pmatrix} \,; \quad \Theta_{\alpha i} \equiv V_{\alpha N_i} = \frac{v(Y_\nu)_{\alpha i}}{\sqrt{2}M_i} \,.
\end{align}
%%%%%%%%%%%%%%%%%%%%%%%%%%%%%
The physical states are three massless Weyl fermions ($\nu'$) and massive Dirac fermions ($N'$).
%%%%%%%%%%%%%%%%%%%%%%%%%%%%%%
This minimal framework must be extended to account for the observed light-neutrino masses. One possibility is to introduce small LNV interactions with $|\Delta L|=2$, such as the small Majorana mass term, $\mu\bar{S}S^c$~(inverse seesaw~\cite{Mohapatra:1986bd}) or $m_L\bar{\nu}S^c$~(linear seesaw~\cite{Akhmedov:1995ip,Akhmedov:1995vm}). In this case, each heavy Dirac state splits into a pair of nearly degenerate Majorana fermions, giving rise to pseudo-Dirac HNLs. However, as demonstrated in Ref.~\cite{Anamiati:2016uxp}, in this regime, LNV processes are strongly suppressed, such that the heavy states decay predominantly through LNC channels, while their production cross section approaches that expected for a Dirac neutrino. Hence,
the phenomenology of Dirac and pseudo-Dirac HNLs with active--sterile mixing is otherwise similar to that of Majorana HNLs, except that LNV signatures are absent in the exact Dirac limit and suppressed in the pseudo-Dirac regime~\cite{Anamiati:2016uxp,Drewes:2019byd,Abada:2022wvh}.
%%%%%%%%%%%%%%%%%%%%%%%%%%%%%%%%%%%%
\section{EFT operators and meson decays}
\label{sec:meson}
%%%%%%%%%%%%%%%%%%%%%%%%%%%%%%%%%
In this work, we consider the dimension six four fermion operators. For Dirac HNL case, these are listed in Table.~\ref{tab:fermionic}. As, we are interested in the HNL production from meson decays, not all of the operators listed in Table.~\ref{tab:fermionic} are equally important. The contact operators triggering meson decays into HNLs involve a pair of quarks and a leptonic part, which in turn contains either (i) a pair of HNLs, or (ii) $N$ and $\nu$ , or else (iii) $N$ and a charged lepton. Some of these possibility has been investigated in detail in Refs.~\cite{DeVries:2020jbs,Cottin:2021lzz,Beltran:2021hpq,Beltran:2022ast,Beltran:2025ilg}, in the
context of long-lived HNLs that could be produced in meson decays at the far-forward and beam dump experiments. In this work, we focus on the operators with pair of HNL, focusing on the case of a single HNL generation. In the group of pair-$N$ operators, $\mathcal{O}_{uN}$, $\mathcal{O}_{dN}$ and $\mathcal{O}_{QN}$ will contribute to the meson decay~\footnote{At dimension six, an additional operator involving two HNL fields can be constructed as
$\mathcal{O}_{HN}
=
\left(H^\dagger i\overleftrightarrow{D_\mu}H\right)
\left(\overline{N}\gamma^\mu N\right).
$
Since this operator generally gives only a subleading contribution to HNL production through meson decays, it is not relevant to the present analysis and is therefore omitted from Table~\ref{tab:fermionic}.
}. Thus, in this work we focus on the $N_R$ pair operators with quarks. 
The corresponding $d=6$ Lagrangian reads~%
\begin{equation}
 \mathcal{L}_{6} \supset \frac{1}{\Lambda^2} \left(
 c_{dN} \mathcal{O}_{dN} + c_{uN} \mathcal{O}_{uN} + c_{QN} \mathcal{O}_{QN}
 \right)
 = \frac{1}{\Lambda^2} \left(
 c_{dN}^{ij} \mathcal{O}_{dN}^{ij} + 
 c_{uN}^{ij} \mathcal{O}_{uN}^{ij} + 
 c_{QN}^{ij} \mathcal{O}_{QN}^{ij}
 \right),
 \label{eq:EFT}
\end{equation}
where a sum over the flavour indices $i,j = 1,2,3$ of quarks.
%In the numerical part of the work, for simplicity we consider only one generation of the HNLs and only the first generation of the quarks, and thus, simplify the notation to $c_{dN}^{11}$, $c_{uN}^{11}$, $c_{QN}^{11}$ and $\mathcal{O}_{dN}^{11}$, $\mathcal{O}_{uN}^{11}$, $\mathcal{O}_{QN}^{11}$, with $``11"$ referring to the first quark generation.
%%%%%%%%%%%%%%%%%%%%%%%%%%%%%%%%%%%
\begin{table}[]
\renewcommand{\arraystretch}{1.5}
\centering
\begin{tabular}{|c | c c|}
\hline
% 4F: RRRR
\multirow{3}{*}{RRRR}&\multicolumn{2}{c|}{$\mathcal{O}_{NN}=(\overline{N}\gamma_\mu N)(\overline{N}\gamma^\mu N)$} \\
&${\cal O}_{eN}=(\overline{e_R}\gamma_\mu e_R)(\overline{N}\gamma^\mu N)$&${\cal O}_{uN}=(\overline{u_R}\gamma_\mu u_R)(\overline{N}\gamma^\mu N)$\\
&${\cal O}_{dN}=(\overline{d_R}\gamma_\mu d_R)(\overline{N}\gamma^\mu N)$&${\cal O}_{duNe} (+\text{h.c.})=(\overline{d_R}\gamma_\mu u_R)(\overline{N}\gamma^\mu e_R)$\\
\hline
% 4F: LLRR
LLRR&${\cal O}_{LN}=(\overline{L}\gamma_\mu L)(\overline{N}\gamma^\mu N)$&${\cal O}_{QN}=(\overline{Q}\gamma_\mu Q)(\overline{N}\gamma^\mu N)$\\
\hline
% 4F: LRLR
\multirow{2}{*}{LRLR}&${\cal O}_{LNLe} (+\text{h.c.})=(\overline{L} N)\epsilon (\overline{L}e_R)$&${\cal O}_{LNQd} (+\text{h.c.})=(\overline{L} N)\epsilon (\overline{Q} d_R)$\\
& \multicolumn{2}{c|}{${\cal O}_{LdQN} (+\text{h.c.})=(\overline{L}d_R)\epsilon (\overline{Q} N)$} \\
\hline
% $4F: LRRL
LRRL & \multicolumn{2}{c|}{${\cal O}_{QuNL} (+\text{h.c.})=(\overline{Q}u_R)(\overline{N}L)$  }  \\
\hline
\end{tabular}
\caption{List of all possible four-fermion lepton and baryon number conserving operators in the presence of $N$ that appear in dimension six construction. Flavor indices are not shown explicitly. Similar operators exists in presence of $S$.  Note that operators $\mathcal{O}_{LNQd}$ and $\mathcal{O}_{LdQN}$ have the same particle content, but represent independent Lorentz structures.}
\label{tab:fermionic}
\end{table}
%%%%%%%%%%%%%%%%%%%%%%%%%%%%%%%%%%%%%%
%%%%%%%%%%%%%%%%%%
\begin{table}[]
\centering
\renewcommand{\arraystretch}{1.25}
\begin{tabular}{|c| c| c|}
\hline
Quark transition & Parent meson & Decay modes \\
\hline
\hline
\multirow{4}{*}{$c\to u$}
 & $D^0$   & $D^0\to N\overline{N}$ \\
 & $D^0$   & $D^0\to \{\pi^0,\eta,\eta',\rho^0,\omega\} N\overline{N}$ \\
 & $D^+$   & $D^+\to \{\pi^+,\rho^+\}N\overline{N}$ \\
 & $D_s^+$ & $D_s^+\to \{K^+,K^{*+}\} N\overline{N}$ \\
\hline
\multirow{4}{*}{$b\to d$}
 & $B^0$   & $B^0\to N\overline{N}$ \\
 & $B^0$   & $B^0\to \{\pi^0,\eta,\eta',\rho^0,\omega\} N\overline{N}$ \\
 & $B^+$   & $B^+\to \{\pi^+,\rho^+\}N\overline{N}$ \\
 & $B_s^0$ & $B_s^0\to \{\overline{K}^{0},\overline{K}^{*0}\}N\overline{N}$ \\
\hline
\multirow{4}{*}{$b\to s$}
 & $B_s^0$ & $B_s^0\to N\overline{N}$ \\
 & $B^0$   & $B^0\to \{K^0, K^{*0}\} N\overline{N}$ \\
 & $B^+$   & $B^+\to \{K^+,K^{*+}\}N\overline{N}$ \\
 & $B_s^0$ & $B_s^0\to \{\eta,\eta',\phi\} N\overline{N}$ \\
\hline
\end{tabular}
\caption{Meson decay channels producing an HNL pair in presence of $\mathcal{O}_{uN}^{12}$~(first row), $\mathcal{O}_{dN}^{31}$~(middle row) and $\mathcal{O}_{dN}^{32}$~(last row).}
\label{tab:meson_HNL_pair}
\end{table}
%%%%%%%%%%%%%%%%%%%%%%%%%%%%
\begin{figure}[!h]
\centering
\includegraphics[width=0.45\linewidth]{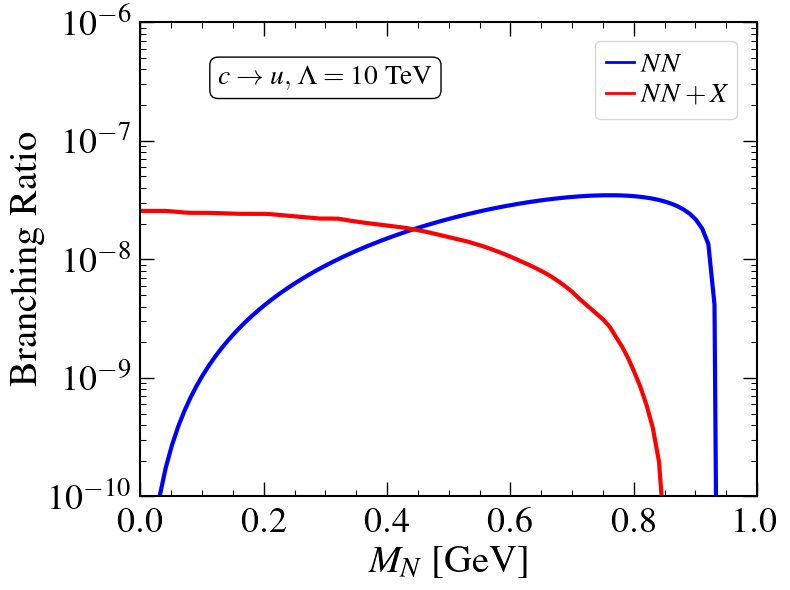}
\includegraphics[width=0.45\linewidth]{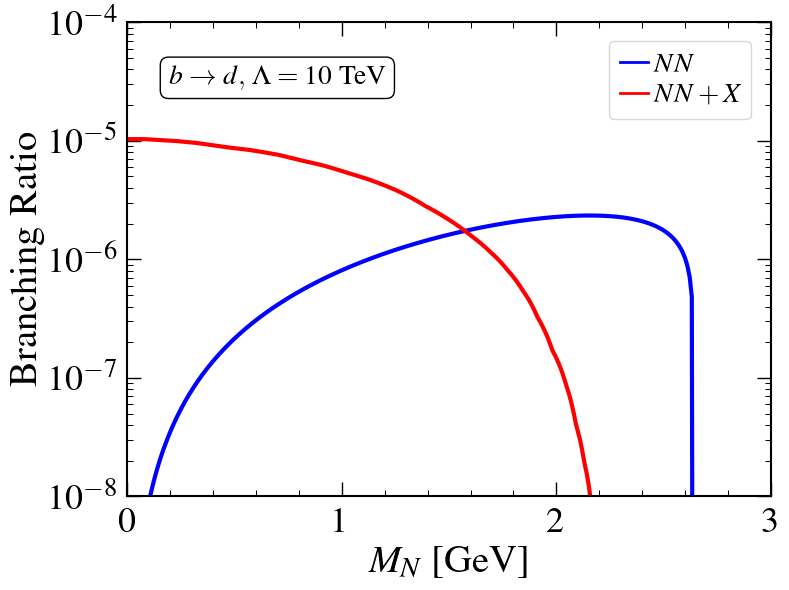}
\includegraphics[width=0.45\linewidth]{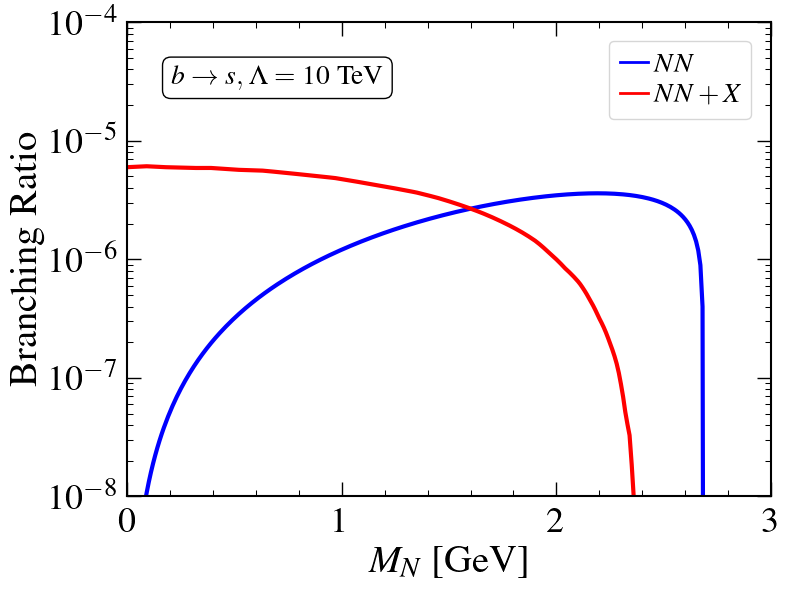}
\caption{Branching ratios of $D$ and $B$ meson decays triggered by the LNC pair-$N_R$ operators $\mathcal{O}_{uN}^{12}$~(top left), $\mathcal{O}_{dN}^{31}$~(top right) and $\mathcal{O}_{dN}^{32}$~(bottom). Blue and red lines stand for two and three-body meson decays with $X=P'$ or $V$.}
\label{fig:BR}
\end{figure}
%%%%%%%%%%%%%%%%%%%%%
\par In the theoretical framework considered in this paper, the HNL production at the LHC
may occur either via the effective operators with a pair of HNLs or the mixing with the
electron neutrinos (mediated by $W$ and $Z$ bosons). However here we are interested in pair production of HNLs in meson decays via EFT operators. In presence of EFT operators as given in Eq.~\ref{eq:EFT}, we can have following two-body~(leptonic) and three-body~(semi-leptonic) meson decays, $P(\overline{q_i}q_j)\to N\overline{N}$ and $P(\overline{q}q_j)\to P'/V(\overline{q}q_i)+N\overline{N}$, respectively~\footnote{Vector mesons with the same quark content have much smaller lifetimes, and thus, their contribution in the HNL production is sub-dominant~\cite{DeVries:2020jbs}.}. In appendix~\ref{app:meson-decay}, we provide the formulae for the two/three-body decay widths. Given their large production rates at hadron collider, we focus on pseudoscalar mesons containing either a $c$ or $b$ quark and a light quark, i.e. $D$- and $B$-mesons. Hence, we focus on following three possibilities: $c\to u$~(for the up-quark sector) and $b\to d$, $b\to s$~(for the down-quark sector). For simplicity we assume that $c\to u$, $b\to d$ and $b\to s$ transition happens via the operators $\mathcal{O}_{uN}^{12}$, $\mathcal{O}_{dN}^{31}$ and $\mathcal{O}_{dN}^{32}$, respectively. In Table.~\ref{tab:meson_HNL_pair}, we listed relevant two and three-body meson decays which can produce pair of HNL.
%%%%%%%%%%%%%%%%%%%%%%%%%%%%%%
\par We have computed the corresponding partial decay widths and in Fig.~\ref{fig:BR} we show the branching ratios of $D$ and $B$-meson decays induced by the LNC pair-$N_R$ operators $\mathcal{O}_{uN}^{12}$~(top left), $\mathcal{O}_{dN}^{31}$~(top right) and $\mathcal{O}_{dN}^{32}$~(bottom), respectively. In each panel, blue stands for two-body mode, whereas red line stands for total three-body meson decays. We see that the two-body decay dominates over the three-body when $M_N>0.4$~GeV for $\mathcal{O}_{uN}^{12}$, $M_N>1.5$~GeV for $\mathcal{O}_{dN}^{31}$ and $M_N>1.6$~GeV for $\mathcal{O}_{dN}^{32}$, respectively. Note that for these we choose the cut-off scale $\Lambda=10$ TeV and the corresponding Wilson coefficients to be one. If the HNL decay products remain undetected, the processes considered here contribute to invisible meson decays, \(P\to \mathrm{inv}\), or to semivisible channels of the form \(P\to P^0/V+\nu\bar{\nu}\). The current upper limits on the corresponding branching ratios are given in Ref.~\cite{ParticleDataGroup:2026mpi}. We find that the most stringent constraints arise from \(D^0/B^0\to \mathrm{inv}\), \(B^0\to\pi^0+\nu\bar{\nu}\), and \(B^+\to K^++\nu\bar{\nu}\). These constraints can be simultaneously satisfied by requiring the cutoff scale to be \(\Lambda>8~\mathrm{TeV}\).
%%%%%%%%%%%%%%%%%%%%%%%
%%%%%%%%%%%%%%%%%%%%%%%%%%%
\section{HNL decay rates}
\label{sec:HNL_decay}
%%%%%%%%%%%%%%%%%%%%%%%%%%%%%%%%%%
In the scenario considered here, the experimentally accessible HNL is the lightest sterile state. Its decays therefore proceed through active--sterile mixing, since the HNL-pair EFT operators do not induce its decay into purely SM final states. We focus on HNL masses well below the SM gauge-boson masses, for which mixing-induced decays are mediated by off-shell $W$ and $Z$ bosons. These interactions give rise to three-body leptonic decays and, where kinematically allowed, decays involving hadrons. The corresponding partial widths have been extensively studied in the literature, and we do not reproduce their analytical expressions here.
For a Dirac HNL, the total decay width is
%%%%%%%%%%%%%%%%%%%%%%%%%%%%
\begin{equation}
\begin{aligned}
\Gamma_N ={}&
\sum_{\ell_1\neq\ell_2}
\left[
\Gamma(N\to\ell_1^-\ell_2^+\nu_{\ell_2})
+
\Gamma(N\to\nu_{\ell_1}\ell_2^-\ell_2^+)
\right]
\\
&+
\sum_{\ell}
\Gamma(N\to\nu_{\ell}\ell^-\ell^+)
+
\sum_{\ell}
\Gamma(N\to\nu_{\ell}\nu\bar{\nu})
+
\Gamma_N^{\rm semileptonic},
\label{decn}
\end{aligned}
\end{equation}
%%%%%%%%%%%%%%%%%%%%%%%%%%%%
where $\ell_1,\ell_2,\ell_3=e,\mu,\tau$, and the three-neutrino contribution includes a sum over the flavours of the $\nu\bar{\nu}$ pair. The semileptonic contribution is evaluated using the kinematically accessible pseudoscalar and vector mesons:
%%%%%%%%%%%%%%%%%%%%%%%%%%%%
\begin{equation}
\begin{aligned}
\Gamma_N^{\rm semileptonic} ={}&
\sum_{\ell,P^+}\Gamma(N\to\ell^-P^+)
+
\sum_{\ell,V^+}\Gamma(N\to\ell^-V^+)
\\
&+
\sum_{\ell,P^0}\Gamma(N\to\nu_{\ell}P^0)
+
\sum_{\ell,V^0}\Gamma(N\to\nu_{\ell}V^0).
\end{aligned}
\end{equation}
%%%%%%%%%%%%%%%%%%%%%%%%%%%%
Here, $P$ and $V$ denote pseudoscalar and vector mesons, respectively. We use the analytical expressions for the partial widths given in Ref.~\cite{Bolton:2025tqw}, with meson masses and decay constants taken from Refs.~\cite{ParticleDataGroup:2010dbb,CLEO:2005jsh,MILC:2002lnl,Ivanov:2006ni,Feldmann:1999uf,Cvetic:2004qg,Ebert:2006hj}.
Since only one HNL is assumed to lie within the experimentally accessible mass range, we do not explicitly fit neutrino oscillation data. Instead, we treat the HNL mass $M_N$, the active--sterile mixing parameters $|V_{\alpha N}|^2$, and the effective operator coefficients $c_{\mathcal O}/\Lambda^2$ as independent parameters. For simplicity, we consider mixing exclusively with the electron flavour, $V_{eN}\neq 0,
V_{\mu N}=V_{\tau N}=0$.
Under this assumption, the total decay width depends on $M_N$ and $|V_{eN}|^2$, while the EFT coefficients control HNL production.

For sufficiently small mixing, the HNL is long-lived and can travel macroscopic distances before decaying, giving rise to displaced decay signatures. Increasing the mass or mixing generally shortens its lifetime and can lead to prompt decays. In this work, we focus on the long-lived regime relevant to FASER2, FACET, SHiP, and FPF@FCC. These experiments probe different laboratory-frame decay lengths because of their different distances from the production point: SHiP, FACET, and FASER2 have baselines ranging from tens to several hundred metres, whereas the FPF@FCC configurations are located approximately $1.5$--$2.0$~km downstream. The sensitivity also depends on the HNL boost and the probability of decay within the detector volume.

For the electron-only mixing scenario, all partial widths scale with $|V_{eN}|^2$, so the branching ratios depend primarily on $M_N$. We define the inclusive visible branching ratio as $\mathrm{BR}(N\to\mathrm{visible})
=
1-\mathrm{BR}(N\to3\nu)$,
where the three-neutrino final state is invisible. At sufficiently low HNL masses, this invisible channel dominates. As additional charged-lepton and hadronic channels become kinematically accessible, including those involving light mesons such as $\pi$ and $\eta$, the visible branching ratio increases and can exceed $90\%$~\cite{A:2025ygb}. Visible decays therefore provide a substantial contribution to the potential signal yield over much of the mass range considered. The observable event rate further depends on the HNL decay probability, the detector acceptance, and the reconstruction efficiency for the relevant final states.
%%%%%%%%%%%%%%%%%%%%%%%%%
\section{Experiments and details of simulation}
\label{sec:simulation}
%%%%%%%%%%%%%%%%%%
Beam-dump and far-forward experiments provide a particularly suitable environment to search for feebly interacting LLPs. Owing to their suppressed interactions with the SM particles, LLPs can travel macroscopic distances before decaying, allowing them to traverse the shielding material and reach a downstream detector. At the same time, the large number of particles produced in high-intensity proton collisions can compensate for the small production rates associated with such weakly coupled states. These experiments can therefore probe regions of parameter space that are difficult to access through conventional prompt searches.
%%%%%%%%%%%%%%%%%%%%%%%%%%
\par The sensitivity of a given experiment is determined by an interplay between the production rate of the LLP, its lifetime and boost, and the geometrical configuration of the detector. In particular, an LLP produced at the interaction point must propagate sufficiently far to reach the detector while still having an appreciable probability of decaying inside its fiducial volume. In addition, its trajectory and visible decay products must satisfy the geometrical acceptance and event-selection requirements of the experiment. The expected number of observable events can consequently be factorized into the LLP production rate, the probability of decay within the fiducial volume, the branching fraction into the relevant visible final states, and the corresponding detector acceptance and efficiency. In our current study, we are interested in forward detectors such as FASER2, FACET and FPF@FCC, as well as proton beam dump setup such as SHiP. The number of signal events for each of these experiments can be formulated as:
%%%%%%%%%%%%%%%%%%%%%%%%%%%
\begin{eqnarray}
N_S = \sum_{i} 2 \cdot  N_{M_i} \cdot \text{BR}(M_i \to N \overline{N}+\text{anything}) \cdot \langle P_{\rm decay} \epsilon_{\rm det} \rangle \cdot \text{BR}(N \to \text{vis.}),\label{eq:Ns}
\end{eqnarray}
where the factor 2 accounts for the fact that each decay of the mesons into the HNLs via the EFT operator includes two HNLs, $N_{M_i}$ is the number of the mother mesons, and $\text{BR}(N \to \text{vis.})$ denotes the decay branching ratio of the HNLs into visible final states. 
For our study, we consider only the decay channel into three active neutrinos to be invisible. In principle, the computation of the decay probability of each HNL can be summarized with a simple formula, if the flight path of the HNL traverses the detector:
%%%%%%%%%%%%%%%%%%%%%%%%%%%%%%%%
\begin{eqnarray}\label{eq:decayprob}
P_{\rm decay} = \text{e}^{-L_1/L_N} -  \text{e}^{-L_2/L_N},
\end{eqnarray}
%%%%%%%%%%%%%%%%%%%%%%%
where $L_N=\beta_N\gamma_N c\tau_N$  and $L_1$ and $L_2$ label respectively the distance from the interaction point~(IP) (where we assume the heavy mesons decay into the HNLs instantly) to the incoming and outgoing sides of the detector. $L_2 = L_1 + L_D $, given $L_D$ being the detector length. HNLs are typically emitted from the source with a non-zero angle~($\theta$) with respect to the beam axis, only a certain fraction of them reach the detector~($\theta\le \alpha$), and even a smaller fraction decay within the detector. Hence, the fraction of produced HNLs which decays inside the detector volume is evaluated by the acceptance factor $ \langle P_{\rm decay} \epsilon_{\rm det} \rangle$. Eq.~\ref{eq:Ns} works directly for SHiP, FASER2, FACET, FPF@FCC, etc.; what changes between experiments is primarily the production spectrum, geometry/acceptance, $L_{1}$, $L_2$, and reconstruction efficiency. The benchmark detector geometries and integrated luminosities or proton exposures are summarized in Table.~\ref{tab:forward_llp_detectors}. We briefly describe the individual experimental setups below. 
%%%%%%%%%%%%%%%%%%%%%%%%%
\begin{table}[t]
\centering
\renewcommand{\arraystretch}{1.25}
\begin{tabular}{lcccc}
\hline\hline
Experiment
& Distance~($L_1$)
& Decay length
& Geometry
& Luminosity \\
\hline
FASER
& $480~\mathrm{m}$
& $1.5~\mathrm{m}$
& $R=0.1~\mathrm{m}$
& $300~\mathrm{fb}^{-1}$ \\

FASER2
& $620~\mathrm{m}$
& $10~\mathrm{m}$
& $3~\mathrm{m}\times1~\mathrm{m}$
& $3~\mathrm{ab}^{-1}$ \\

FACET
& $\simeq100~\mathrm{m}$
& $18~\mathrm{m}$
& $R=0.5~\mathrm{m}$
& $3~\mathrm{ab}^{-1}$ \\

FPF1 (FCC-hh)
& $1.5~\mathrm{km}$
& $50~\mathrm{m}$
& $5~\mathrm{m}\times5~\mathrm{m}$
& $30~\mathrm{ab}^{-1}$ \\

FPF2 (FCC-hh)
& $1.5~\mathrm{km}$
& $400~\mathrm{m}$
& $20~\mathrm{m}\times20~\mathrm{m}$
& $30~\mathrm{ab}^{-1}$ \\

SHiP
& $\simeq 33~\mathrm{m}$
& $50~\mathrm{m}$
& $4\text{ m}\times \text{ 6 m}$
& $6\times10^{20}~\mathrm{PoT}$ \\ \hline
\end{tabular}
\caption{Benchmark parameters for forward LLP decay searches. The distance is measured from the interaction point, or from the proton target for SHiP, to the start of the decay volume. The transverse dimensions describe the decay volume unless otherwise
specified. FPF1 and FPF2 denote the FCC-hh configurations.}
\label{tab:forward_llp_detectors}
\end{table}
%%%%%%%%%%%%%%%%%%%%%%%%%%
\paragraph{\textbf{FASER2}:} FASER~\cite{Kling:2018wct,FASER:2018ceo,FASER:2018bac,
Feng:2017vli,FASER:2018eoc} searches for light, weakly interacting particles produced at the ATLAS IP. It is located
$480~\mathrm{m}$ downstream of the IP, on the beam collision axis in the TI12 service tunnel~\cite{FASER:2022hcn}. The limited underground space constrains its decay volume, motivating the larger FASER2 detector proposed for the HL-LHC~\cite{Feng:2017uoz,Feng:2022inv}. Proposed locations include a purpose-built facility approximately $617$--$682~\mathrm{m}$
west of the IP and alcoves near the UJ12 cavern approximately
$480$--$521~\mathrm{m}$ east of the IP~\cite{Feng:2022inv,Anchordoqui:2021ghd}. We adopt the purpose-built configuration and approximate its fiducial decay
volume by a cylinder with
\begin{equation}
  L=620~\mathrm{m},\qquad
  \Delta=10~\mathrm{m},\qquad
  R=1~\mathrm{m},\qquad
  \mathcal{L}=3~\mathrm{ab}^{-1},
\end{equation}
where $L$ is the distance from the IP to the detector, while $\Delta$ and
$R$ denote the decay-volume length and radius. We assume
$\sqrt{s}=14~\mathrm{TeV}$. A representative detector layout places
tracking stations, a magnetic spectrometer, calorimeters, and a muon
identification system downstream of the decay volume. To get the $D/B$ meson spectra at FASER2 we have used forward corrected meson spectra from $\mathrm{FORESEE}$ ~\cite{Kling:2021fwx} package. HNLs produced in
forward $D$- and $B$-meson decays may traverse the intervening shielding
and decay visibly within this volume. Their trajectories must intersect
the fiducial region; reconstructing a decay further requires its charged
products to enter the downstream tracking and spectrometer acceptance.
The cylindrical geometry corresponds to a characteristic angular reach
$R/L\simeq1.6~\mathrm{mrad}$ and therefore selects only HNLs within this angular coverage. We also impose a trigger requirement of $p_{\rm HNL}>100~{\rm GeV}$, which is typically satisfied by highly boosted HNLs travelling close to the beam collision axis and reaching the FASER2 detector before decaying~\footnote{At the LHC, light particles are typically produced with transverse momenta comparable to their masses, $p_T\sim m_{\rm LLP}$. Consequently, LLPs entering FASER's narrow angular acceptance, $\theta\lesssim 1~{\rm mrad}$, generally have energies of order a TeV~\cite{FASER:2018eoc}.}.
%%%%%%%%%%%%%%%%%%%%%%%%%%%%%%
\paragraph{\textbf{FACET}:}
%%%%%%%%%%%%%%%%%%%%%%%%%%%%%%
FACET (Forward-Aperture CMS ExTension) is a proposed detector for LLPs produced in the very forward direction of the CMS interaction point (IP5) at the HL-LHC~\cite{Cerci:2021nlb}. In the
original design, particles pass through approximately $50~\mathrm{m}$
of magnetized iron before reaching an enlarged section of the LHC beam
pipe that serves as an evacuated decay volume. This cylindrical region
has a radius of $0.5~\mathrm{m}$ and extends from approximately
$z=101$ to $119~\mathrm{m}$ downstream of IP5. Its visible decay
products would be measured using downstream tracking, precision timing,
calorimetry, and muon-identification systems~\cite{Cerci:2021nlb}.
For the HNL sensitivity estimate, we require the HNL to reach and decay
inside the fiducial volume, with its visible decay products entering
the downstream detectors. The proposed downstream acceptance is annular,
with inner and outer radii of approximately $0.18$ and $0.50~\mathrm{m}$,
respectively. It is commonly approximated by
$1.5\lesssim\theta_N\lesssim4~\mathrm{mrad}$, corresponding to
$6.2\lesssim\eta_N\lesssim7.2$
~\cite{Hacisahinoglu:2025xrs}.
An upstream charged-particle veto, displaced-vertex reconstruction,
and proposed timing resolution of order $30~\mathrm{ps}$ provide means
of suppressing backgrounds~\cite{Cerci:2021nlb}.
%%%%%%%%%%%%%%%%%%%%%%%%%%%%%%%
\paragraph{\textbf{FPF@FCC}:}
A proposed extension of the far-forward long-lived-particle program is
the Forward Physics Facility at the Future Circular Collider
(FPF@FCC)~\cite{FCC:2018byv,FCC:2018vvp,Aleksa:2019pvl,
MammenAbraham:2024gun}. At the FCC-hh, $pp$ collisions at
$\sqrt{s}=100~\mathrm{TeV}$ with an integrated luminosity of up to
$30~\mathrm{ab}^{-1}$ would produce a large flux of forward
heavy-flavour mesons. Their decays provide a source of HNLs that could
travel to a detector approximately $1.5~\mathrm{km}$ downstream of
the interaction point (IP) and decay visibly. A proposed sweeper magnet
would reduce the forward-muon flux, while approximately $500~\mathrm{m}$
of rock between the IP and detector would provide additional shielding~\cite{MammenAbraham:2024gun}. Following Ref.~\cite{MammenAbraham:2024gun}, we consider two benchmark
detectors centred on the collision-axis line of sight:
\begin{align}
\mathrm{FPF1}\;(\mathrm{FCC\!-\!LLP1}):\,\,
5~\mathrm{m}\times5~\mathrm{m}\times50~\mathrm{m},\quad
\mathrm{FPF2}\;(\mathrm{FCC\!-\!LLP2}):\,\,
20~\mathrm{m}\times20~\mathrm{m}\times400~\mathrm{m}.
\label{eq:fpf_fcc_geometry}
\end{align}
In both cases the upstream face is located $L_1=1.5~\mathrm{km}$
from the IP, and we assume $\mathcal{L}=30~\mathrm{ab}^{-1}$.
FPF1, which we take as the baseline configuration, has a
$50~\mathrm{m}$ decay length comparable to that of
SHiP~\cite{SHiP:2018xqw,SHiP:2025ows}; FPF2 provides a larger
fiducial volume. 
For an HNL produced at the IP, the transverse position at the detector
entrance must satisfy $|x(L_1)|,|y(L_1)|<L_3$, where $L_3=2.5~\mathrm{m}$
for FPF1 and $10~\mathrm{m}$ for FPF2. The corresponding characteristic
angular scales are $L_3/L_1\simeq1.7~\mathrm{mrad}$ and
$6.7~\mathrm{mrad}$, respectively. An accepted HNL must survive to
the detector and decay within its fiducial volume; its visible decay
products must also enter the downstream tracking and spectrometer
acceptance. Similar to FASER2, we also impose a trigger requirement of $p_{\rm HNL}>100~{\rm GeV}$. The long flight distance
selects HNLs with sufficiently large laboratory-frame decay lengths,
while the narrow angular coverage preferentially accepts highly
boosted HNLs.
%%%%%%%%%%%%%%%%%%%%%%%%%%%%%%%%
%%%%%%%%%%%%%%%%%%%%%%%%%%%%%%%%%%
\paragraph{\textbf{SHiP}:}
SHiP is a high-intensity proton beam-dump experiment at the CERN SPS
designed to search for feebly interacting particles, including long-lived HNLs~\cite{SHiP:2025ows}. A
$400~\mathrm{GeV}$ proton beam strikes a dense target, where charm and
beauty hadrons are produced and can subsequently decay into HNLs. 
Downstream of the target, a hadron stopper and an active magnetic
muon shield suppress ordinary particles entering the detector.
In the ECN3 configuration, the fiducial decay region begins
approximately $33~\mathrm{m}$ from the target centre and extends for
about $50~\mathrm{m}$~\cite{ship}. Its transverse dimensions
increase from approximately
$1.0~\mathrm{m}\times2.7~\mathrm{m}$ at the upstream face to
$4~\mathrm{m}\times6~\mathrm{m}$ at the downstream face. The decay
region is surrounded by background taggers and is followed by
tracking stations, a magnetic spectrometer, timing and particle
identification systems, and calorimeters~\cite{ship,SHiP:2025ows}. The current SHiP concept uses
a lightweight helium-filled decay region to limit interactions in
the fiducial volume~\cite{SHiP:2025ows}.
In our simplified sensitivity
estimate, we approximate the widening fiducial region by a cone
and impose $\theta_N<\theta_{\rm max}=2.8\times10^{-2}$ relative
to the beam axis, following Ref.~\cite{Barducci:2024nvd}.
This angular cut approximates the HNL's geometric acceptance;
the requirement on its visible decay products must be applied
separately. The muon shield, surrounding vetoes, low-material
decay region, and vertex and timing selections are designed to
suppress muon and neutrino-induced backgrounds~\cite{SHiP:2025ows}.\\
%%%%%%%%%%%%%%%%%%%%%%%%%%%%%%%
%%%%%%%%%%%%%%%%%%%%%%%%%%%%%
%%%%%%%%%%%%%%%%%%%%%%%%%%%%%%%%
\begin{table}[t]
    \centering
    \renewcommand{\arraystretch}{1.25}
    \begin{tabular}{lccc}
        \hline\hline
        Meson species
        & SHiP
        & HL-LHC
        & FCC-hh \\
        & $6\times10^{20}$ POT
        & $\sqrt{s}=14~\mathrm{TeV}$,
          $\mathcal{L}=3~\mathrm{ab}^{-1}$
        & $\sqrt{s}=100~\mathrm{TeV}$,
          $\mathcal{L}=30~\mathrm{ab}^{-1}$ \\
        \hline
        $D^0$
        & $1.29\times10^{18}$
        & $4.14\times10^{16}$
        & $1.03\times10^{18}$ \\

        $D^\pm$
        & $4.2\times10^{17}$
        & $2.12\times10^{16}$
        & $5.3\times10^{17}$ \\

        $D_s^\pm$
        & $1.8\times10^{17}$
        & $7.1\times10^{15}$
        & $1.77\times10^{17}$ \\

        $B^0,B^\pm$
        & $8.1\times10^{13}$
        & $2.78\times10^{15}$
        & $1.02\times10^{17}$ \\

        $B_s^0$
        & $2.16\times10^{13}$
        & $2.73\times10^{14}$
        & $1.01\times10^{16}$ \\
        \hline\hline
    \end{tabular}
\caption{Estimated numbers of heavy mesons produced at SHiP, the HL-LHC, and FCC-hh. These are inclusive
production-level estimates before geometrical acceptance, decay probabilities, and detector efficiencies are applied.}
    \label{tab:heavy_meson_yields}    
\end{table}
%%%%%%%%%%%%%%%%%%%%%%%%%%%%%%%%%%%%%
\par To perform the sensitivity estimate for experiments at LHC/FCC-hh, it is necessary to know the inclusive production 
rate of the heavy-flavor mesons at the HL-LHC/FCC-hh. For this we have used Pythia 8~\cite{Bierlich:2022pfr} to obtain meson spectra resulting from the $pp$ collisions simulation. As discussed in Ref.~\cite{Dercks:2018eua,Bondarenko:2018ptm,deVries:2015mfw, Dreiner:2020qbi}, Pythia 8 has not been extensively validated for the differential production of charm and bottom mesons at large pseudorapidities. We therefore improve its predictions by rescaling the meson-production cross sections in different bins of transverse momentum and pseudorapidity to match the more reliable FONLL results~\cite{Cacciari:1998it,Cacciari:2001td,Cacciari:2012ny,Cacciari:2015fta}. In Table.~\ref{tab:heavy_meson_yields}, we reproduce the inclusive production rates of all the relevant types of heavy mesons at HL-LHC and FCC-hh.
%%%%%%%%%%%%%%%%%%%%%%%%%%%%
\par Since Pythia 8 does not account for post production propagation effects inside the target, we use $\mathrm{FairShip}$~\cite{the_ship_collaboration_2026_21469668} to get the meson spectrum at SHiP. In $\mathrm{FairShip}$ simulations, Pythia 6.4~\cite{Sjostrand:2006za} and $\mathrm{GEANT4}$~\cite{GEANT4:2002zbu} handles the heavy flavour production. Secondary heavy flavour production in cascade interactions of hadrons originated by the initial proton collision~\cite{ruf_2015_30w4z-p7w72} is also taken into account. However $\mathrm{Pythia \ 6.4}$ undermines the charm and beauty production, therefore we impose cascade enhancements as in~\cite{SHiP:2018xqw}.
For our sensitivity estimate, we assume an exposure of
$N_{\rm POT}=6\times10^{20}$ protons on target, corresponding to the
proposed 15-year programme~\cite{SHiP:2025ows,Barducci:2024nvd}. The number of produced mesons of species $\mathcal{M}$ is written as
$N_{\mathcal{M}}=N_{\rm POT}\,f_{\mathcal{M}}$,
where $f_{\mathcal{M}}$ is the average yield of that species per proton
on target, including any secondary production accounted for in the
meson simulation~\cite{Barducci:2024nvd}. Results obtained post cascade corrections are given in Table.~\ref{tab:heavy_meson_yields}.
%%%%%%%%%%%%%%%%%%%%%%%%%%
\par The high particle flux in the far-forward region also poses significant challenges due to SM backgrounds. Two complementary mechanisms are particularly important for their suppression. First, dedicated sweeper magnets can deflect charged particles, particularly high-energy muons, away from the detector acceptance. Second, the rock and additional shielding material separating the interaction point from the detector can effectively attenuate hadronic and electromagnetic backgrounds. In particular, the proposed detector sites such as FASER2/FACET and FPF@FCC  are shielded from the interaction point by $\mathcal{O}(100\text{ m})$ of concrete and rock~\cite{Bohlen:2014buj,Ferrari:2005zk,FASER:2018eoc}. Furthermore, timing and directional information can help discriminate cosmic-ray backgrounds from LLP signals originating at the interaction point. Together, these measures are expected to provide a comparatively low-background environment for LLP searches. Nevertheless, residual backgrounds from muons, neutrinos, and cosmic rays cannot be assumed to vanish entirely and would need to be carefully assessed in a detailed experimental analysis.
For the phenomenological projections presented in the next section, we adopt the idealized assumptions of zero background events and $100\%$ signal detection efficiency. 
 %%%%%%%%%%%%%%%%%%%%%%%
\section{Results and Discussions}
\label{sec:results}
\begin{figure}[!h]
    \centering
    \includegraphics[width=0.95\linewidth]{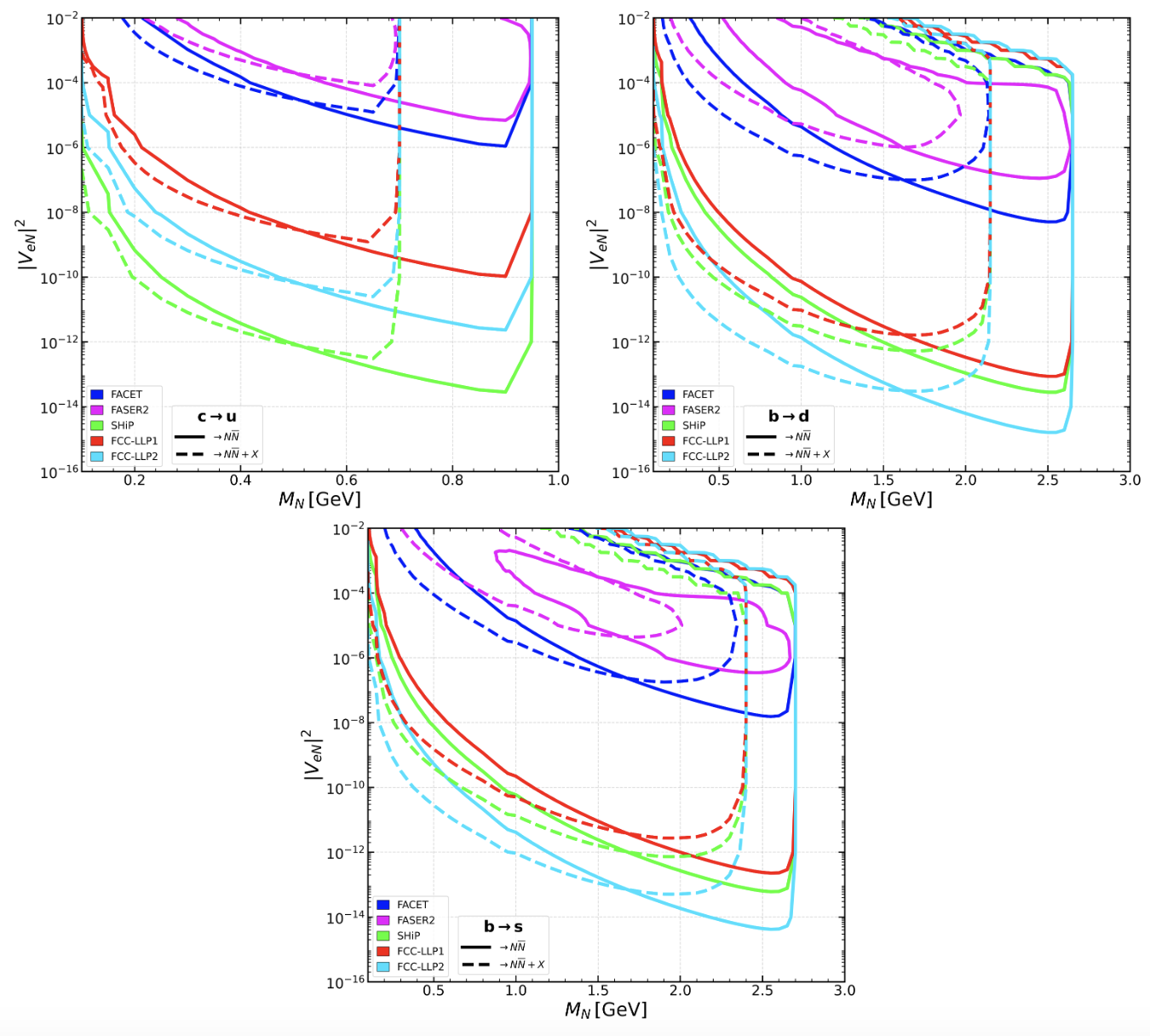}
    \caption{Projected $95\%$ C.L. sensitivity contours in the
    $(M_N,|V_{eN}|^2)$ plane for the three LNC pair-$N_R$ operators
    considered in this work. The relevant dimensionless Wilson
    coefficient is set to unity, and the cut-off scale is fixed at
    $\Lambda=10~\mathrm{TeV}$. Dashed and solid curves show the
    contributions from three-body and two-body meson decays,
    respectively.}
    \label{fig:contour}
\end{figure}

\begin{figure}[!h]
    \centering
    \includegraphics[width=0.95\textwidth]{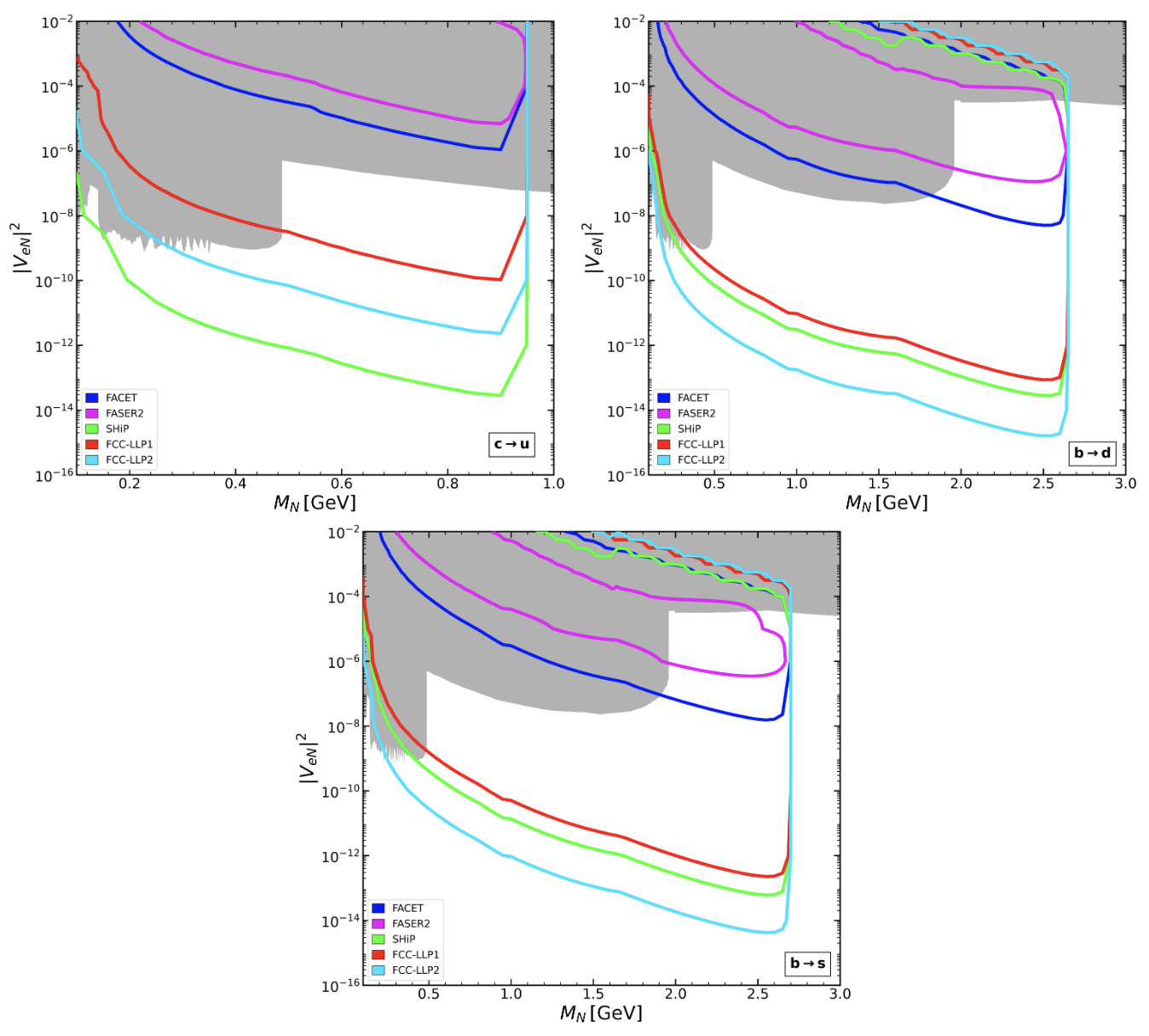}
    \caption{Combined projected $95\%$ C.L. sensitivity contours in
    the $(M_N,|V_{eN}|^2)$ plane, including both two-body and
    three-body meson decays, for $\Lambda=10~\mathrm{TeV}$ and a
    unit dimensionless Wilson coefficient. Existing constraints
    on the electron-flavor mixing are shown in gray.}
    \label{fig:contour-combined}
\end{figure}

%%%%%%%%%%%%%%%%%%%%%%%%%%%
We present the numerical results for the LNC pair-$N_R$ operators
that induce the meson decays discussed in Sec.~\ref{sec:meson}.
We first investigate the sensitivity in the $(M_N,|V_{eN}|^2)$ plane
by varying the HNL mass and active--sterile mixing at a fixed cut-off
scale. The projected sensitivities of the detectors considered in
this work are shown in Figs.~\ref{fig:contour}
and~\ref{fig:contour-combined}.\footnote{In our numerical analysis,
we assume that the HNLs mix predominantly with the electron neutrino
and neglect their mixing with the other active flavors.}
Each contour corresponds to three expected signal events, giving an
approximate $95\%$ confidence-level (C.L.) exclusion reach under the
assumptions of zero background events and unit signal detection
efficiency. We set the relevant dimensionless Wilson coefficient to
unity and fix $\Lambda=10~\mathrm{TeV}$, consistently with the existing
constraints from $D^0/B^0\to\mathrm{inv}$,
$B^0\to\pi^0\nu\bar{\nu}$, and $B^+\to K^+\nu\bar{\nu}$.

%%%%%%%%%%%%%%%%%%%%%%%%%%%%%%%%%%%%%%%%%
%%%%%%%%%%%%%%%%%%%%%%%%%%%%%%%%%%%%%%%%%

The dashed and solid curves in Fig.~\ref{fig:contour} show the
sensitivities obtained separately from three-body and two-body meson
decays into Dirac HNL pairs. At lower HNL masses, the three-body
channels generally provide greater sensitivity to small
$|V_{eN}|^2$, reflecting their larger production branching ratios.
As $M_N$ increases, these channels become increasingly suppressed
by the available phase space, and the two-body channels become
dominant. The latter also extend the accessible mass range toward
$M_N=m_P/2$, where $m_P$ is the parent-meson mass, whereas the
three-body channels are restricted to
$M_N\leq(m_P-m_X)/2$, with $X$ denoting the accompanying meson.
The relative sensitivities additionally depend on the HNL momentum
distributions and the detector acceptance.

The contour shapes reflect the interplay between the HNL lifetime
and the detector geometry. At very small mixing, the HNL decay length
is sufficiently large that most HNLs traverse the detector without
decaying. Increasing $|V_{eN}|^2$ enhances the probability of a decay
within the fiducial volume and therefore increases the signal yield.
At sufficiently large mixing, however, the HNLs predominantly decay
before reaching the detector, producing the upper boundary of the
sensitivity region. The rapid loss of sensitivity near the largest
accessible HNL mass follows from the suppression of meson decays
close to the pair-production threshold.

%%%%%%%%%%%%%%%%%%%%%%%%%%%%%%%%%%%%%%%%%
%%%%%%%%%%%%%%%%%%%%%%%%%%%%%%%%%%%%%%%%%

Fig.~\ref{fig:contour-combined} presents the combined sensitivities
obtained by including both production channels. The $b\to d$ and
$b\to s$ scenarios, shown in the top-right and bottom panels,
respectively, extend to larger $M_N$ than the $c\to u$ scenario
shown in the top-left panel. This difference follows from the larger
masses of the parent $B$ mesons, which permit the production of heavier
HNL pairs. The bottom-meson channels also reach smaller values of
$|V_{eN}|^2$ over substantial portions of the accessible mass range,
despite the larger charm-meson production rates reported in
Table~\ref{tab:heavy_meson_yields}. The parent-meson yield alone
therefore does not determine the sensitivity: the production branching
ratios, HNL kinematics, geometrical acceptance, and decay probability
must all be taken into account.

For the $c\to u$ transition, SHiP provides the strongest sensitivity
to small mixing, followed by FCC-LLP2, FCC-LLP1, FACET, and FASER2.
For the $b\to d$ and $b\to s$ transitions, FCC-LLP2 offers the greatest
reach, followed by SHiP, FCC-LLP1, FACET, and FASER2. These rankings
refer to the lower-mixing boundaries of the sensitivity contours.
Although the inclusive charm-production cross sections at SHiP and
FCC-hh are of a similar order, the observable HNL yield also depends
on the beam exposure, boost distribution, and detector geometry.
In particular, the different HNL boosts affect both the fraction
entering the detector and the probability of decaying within its
fiducial volume. The larger FCC-LLP2 configuration substantially
improves the sensitivity relative to FCC-LLP1 and provides the
strongest reach in the bottom-meson scenarios.

We compare these projections with existing constraints on
$|V_{eN}|^2$, shown in gray in
Fig.~\ref{fig:contour-combined}. For
$M_N\lesssim0.15~\mathrm{GeV}$, peak searches in pion decays at
TRIUMF~\cite{Britton:1992xv} and
PIENU~\cite{PIENU:2017wbj,Bryman:2019bjg,PIENU:2019usb}
provide stringent bounds.
Borexino~\cite{Borexino:2013bot} probes the overlapping interval
$0.003\leq M_N\leq0.016~\mathrm{GeV}$ through
$N\to\nu e^+e^-$. At larger masses, NA62 peak searches in
$K^+\to e^+N$ constrain
$0.15\lesssim M_N\lesssim0.45~\mathrm{GeV}$
~\cite{NA62:2020mcv,NA62:2025csa}, while the T2K near detector
probes $0.4\lesssim M_N\lesssim0.5~\mathrm{GeV}$
~\cite{T2K:2019jwa}. Beam-dump searches at
BEBC~\cite{WA66:1985mfx,Barouki:2022bkt} and
CHARM~\cite{CHARM:1985nku} extend the constraints to approximately
$2.25~\mathrm{GeV}$. At still higher masses, searches by
Belle~\cite{Zhou:2021ylt}, DELPHI~\cite{DELPHI:1996qcc},
ATLAS~\cite{Tastet:2021vwp,ATLAS:2022atq}, and
CMS~\cite{CMS:2022fut} probe the parameter space up to approximately
$16~\mathrm{GeV}$. The projected sensitivities demonstrate that
EFT-induced pair production can provide access to mixing values
substantially below these existing bounds. This improvement arises
because the production rate is controlled by the pair-$N_R$ operator,
whereas the HNL lifetime is governed by the active--sterile mixing.

%%%%%%%%%%%%%%%%%%%%%%%%%%%%%%%%%%%%%%%%%
\begin{figure}[htbp]
    \centering
    \includegraphics[width=0.95\textwidth]{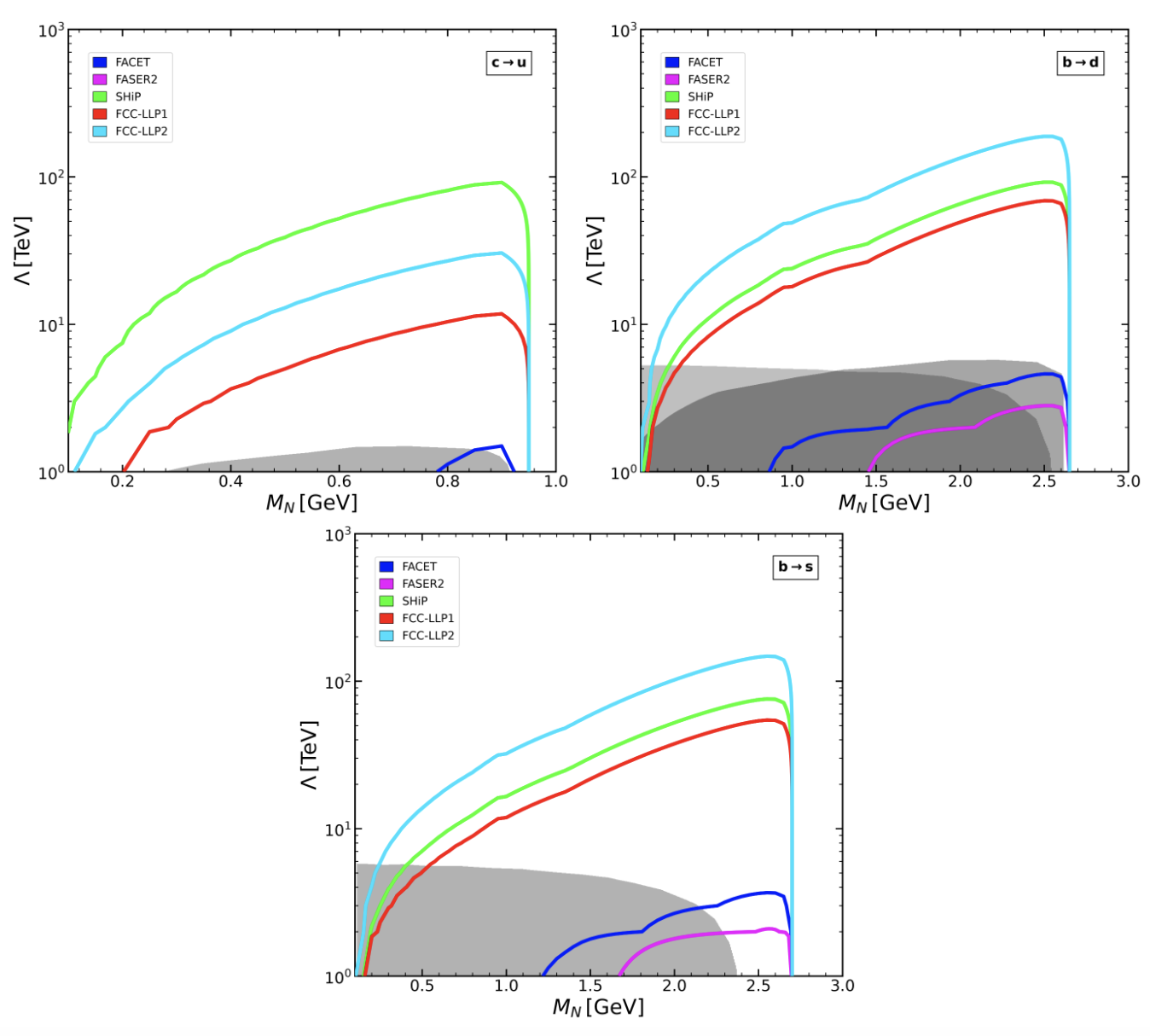}
    \caption{Projected $95\%$ C.L. sensitivity contours in the
    $(M_N,\Lambda)$ plane for $|V_{eN}|^2=10^{-10}$ and a unit
    dimensionless Wilson coefficient, including both two-body and
    three-body meson decays. Existing constraints from invisible
    and semivisible meson decays are shown in gray.}
    \label{fig:lambda}
\end{figure}
%%%%%%%%%%%%%%%%%%%%%%%%%%%%%%%%%%%%%%%%%

We next present the sensitivity in the $(M_N,\Lambda)$ plane in
Fig.~\ref{fig:lambda}. We fix $|V_{eN}|^2=10^{-10}$ and determine,
at each HNL mass, the value of $\Lambda$ for which the expected
signal yield equals three. With the dimensionless Wilson coefficient
set to unity, the production rate scales as $\Lambda^{-4}$, while
the mixing-induced HNL lifetime remains fixed. Each contour therefore
indicates the largest cut-off scale accessible at the corresponding
HNL mass.

The relative performance of the experiments follows the hierarchy
of the small-mixing sensitivities in
Fig.~\ref{fig:contour-combined}. In the charm scenario, SHiP provides
the strongest reach, approaching $\Lambda\sim100~\mathrm{TeV}$,
or equivalently an effective interaction strength
$c_{\mathcal O}/\Lambda^2\sim10^{-4}~\mathrm{TeV}^{-2}$.
In the bottom scenarios, FCC-LLP2 probes scales of order
$(1$--$2)\times10^2~\mathrm{TeV}$, corresponding to effective
interaction strengths of a few times
$10^{-5}~\mathrm{TeV}^{-2}$. Over much of the accessible mass range,
the reach improves with increasing $M_N$ because the larger HNL
decay width enhances the probability of a visible decay within the
detector. Close to the production threshold, however, phase-space
suppression reduces the HNL flux and causes the reach in $\Lambda$
to decrease rapidly.

The gray shaded regions in Fig.~\ref{fig:lambda} represent the
constraints from invisible and semivisible $D$- and $B$-meson
decays discussed earlier. For the $c\to u$ transition, the relevant
constraint comes from $D^0\to\mathrm{inv}$. For $b\to d$,
$B^0\to\pi^0\nu\bar{\nu}$ provides the strongest bound for
$M_N\lesssim1.3~\mathrm{GeV}$, whereas $B^0\to\mathrm{inv}$
dominates at larger masses. For $b\to s$, the leading constraint
comes from the upper limit on the branching ratio of
$B^+\to K^+\nu\bar{\nu}$. SHiP and the FCC-LLP configurations
can probe interaction strengths well below these existing bounds
over broad mass intervals, illustrating the complementarity between
missing-energy searches in meson decays and searches for displaced
visible HNL decays.
%%%%%%%%%%%%%%%%%%%%%%%%%
\section{Conclusion}
\label{sec:conclusion}
%%%%%%%%%%%%%%%%%%%%%%%
In this work, we have investigated the prospects for probing
dimension-six LNC pair-$N_R$ operators within the $\nu$SMEFT
framework through searches for long-lived Dirac HNLs. We focused on
the flavor transitions $c\to u$, $b\to d$, and $b\to s$, which induce
HNL-pair production in two-body and three-body decays of heavy
mesons. We evaluated the sensitivities of FASER2, FACET, SHiP, and
two FCC-LLP benchmark configurations, accounting for the parent-meson
yields, HNL kinematics, geometrical acceptance, and decay probability
within the fiducial volume.

A central feature of this scenario is the separation between HNL
production and decay. While meson decays into HNL pairs are governed
by the effective interaction strength $c_{\mathcal O}/\Lambda^2$,
the subsequent HNL decays are controlled by the active--sterile mixing
$|V_{eN}|^2$. Consequently, appreciable HNL fluxes can be produced
even for very small mixing, where conventional mixing-induced
production becomes inefficient. The meson branching ratios presented
in Fig.~\ref{fig:BR} illustrate the complementary roles of the production
channels: three-body decays generally dominate at lower HNL masses,
whereas two-body decays become increasingly important toward the
kinematic endpoint.

The corresponding contributions to the experimental sensitivity
are shown separately in Fig.~\ref{fig:contour}. Three-body channels
generally provide greater sensitivity to small mixing at lower
masses, while two-body channels extend the reach toward the
pair-production threshold. Combining both contributions gives the
sensitivities shown in Fig.~\ref{fig:contour-combined}. For
$\Lambda=10~\mathrm{TeV}$ and unit dimensionless Wilson coefficients,
SHiP provides the strongest reach in small $|V_{eN}|^2$ for the
$c\to u$ transition, whereas FCC-LLP2 offers the greatest sensitivity
for $b\to d$ and $b\to s$. The bottom-meson channels also extend the
accessible HNL mass range because of the larger parent-meson masses.
These projections reach mixing values substantially below the
existing constraints displayed in
Fig.~\ref{fig:contour-combined}, demonstrating the additional
discovery potential of EFT-induced HNL-pair production.

Fig.~\ref{fig:lambda} expresses the sensitivity in terms of the
cut-off scale for a fixed mixing $|V_{eN}|^2=10^{-10}$. In the charm
scenario, SHiP probes scales approaching
$\Lambda\sim100~\mathrm{TeV}$ for a unit Wilson coefficient,
corresponding to
$c_{\mathcal O}/\Lambda^2\sim10^{-4}~\mathrm{TeV}^{-2}$.
For the bottom scenarios, FCC-LLP2 reaches scales of order
$(1$--$2)\times10^2~\mathrm{TeV}$, corresponding to effective
interaction strengths of a few times
$10^{-5}~\mathrm{TeV}^{-2}$. Over broad mass intervals, these
sensitivities extend beyond the constraints from invisible and
semivisible meson decays shown in the same figure, highlighting
the complementarity between missing-energy measurements and
searches for displaced visible HNL decays.

Our projections assume negligible backgrounds and unit detection
efficiency for visible HNL decays. A detailed treatment of
experimental backgrounds and reconstruction efficiencies will be
required to determine the achievable sensitivity. Within these
assumptions, our results establish heavy-meson decays as a promising
probe of pair-$N_R$ interactions and demonstrate the complementary
roles of high-intensity beam-dump experiments and far-forward
detectors in exploring the heavy-neutrino sector of $\nu$SMEFT.
%%%%%%%%%%%%%%%%%%%%%%%%%%
\acknowledgments
\noindent
We thank the organizers of WHEPP-2025 for providing a stimulating environment in which this work was initiated. The work of SM is supported by KIAS Individual Grants (PG086002) at Korea Institute for Advanced Study. The authors acknowledge the use of ChatGPT (OpenAI) for assistance with language editing and improving the clarity and readability of the manuscript.
%%%%%%%%%%%%%%%%%%%%%%%%
\appendix
%%%%%%%%%%%%%%%%%%%%%%%%%%%%%%%%%%%
\section{Pseudoscalar meson decays into HNLs}
%%%%%%%%%%%%%%%%%%%%%%%%%%%%%%%
\label{app:meson-decay}
This appendix gathers
the necessary formulae to determine two/three-body meson decay rates into HNLs coming from the dimension six operators in Tab.~\ref{tab:fermionic}. The corresponding amplitude for the two-body process $P(\bar{q}_i q_j)\to N\bar{N}$ is given as follows,
%%%%%%%%%%%%%%%%%%%%%%
\begin{align}
\mathcal{M}\left(P \to N\overline{N}\right)  = i \frac{f_P}{2} \left(c_{qN,\, ij}^{V,RR} - c_{qN,\,ij}^{V,LR}\right) \overline{u_N} \slashed{p} P_R v_{N'}\,,  
\end{align}
%%%%%%%%%%%%%%%%%%%%%%%%%%%%%%%%%%%%%%%
where $c_{dN,ij}^{V,RR} = C_{dN}^{ij}$, $c_{uN,ij}^{V,RR} = C_{uN}^{ij}$, $c_{dN,ij}^{V,LR} = V_{ki}^\ast V_{lj} C_{QN}^{kl}$ and $c_{uN,ij}^{V,LR} = C_{QN}^{ij}$. $u_N$ and $v_{N'}$ denote the spinors associated with the HNLs
$N$ and $\overline{N}$, respectively.
We define the meson decay constants through the transition matrix elements 
%%%%%%%%%%%%%%%%%%%%%%
$\langle 0 | \overline{q_i} \gamma^\mu \gamma_5 q_j | P(p) \rangle = i f_P p^\mu$, where $i,j$ denote quark flavor indices, and $ |P(p) \rangle $ is the decaying pseudoscalar meson. We have used the following value for meson decay constant: $f_{D^0}=0.212$~GeV, $f_{B^0}=0.190$~GeV and $f_{B_s^0}=0.230$~GeV.
%%%%%%%%%%%%%%%%%%%%%%%%%%%%%%%%%%%%%
 The corresponding decay rate is given as follows,
%%%%%%%%%%%%%%%%%%%%%%%%%%%%%%%%%%%%%%%%
 \begin{align}
 \Gamma\left(P \to N\overline{N}\right) & = \frac{\left|f_P\right|^2}{32\pi} 
 \left|c_{qN,\, ij}^{V,RR} - c_{qN,\,ij}^{V,LR}\right|^2 m_P \, m_N^2 \, \sqrt{1-\frac{4m_N^2}{m_P^2}}\,, 
\end{align}
%%%%%%%%%%%%%%%%%%%%%%%%%%%%%%%%%%%%%%%%%
The amplitudes of the relevant three-body decays for producing the HNLs via the dimension six operators are
%%%%%%%%%%
\begin{align}
\mathcal{M}\left(P \rightarrow P' N \overline{N} \right) & = \frac{1}{2}  \left(c_{qN,\, ij}^{V,RR} + c_{qN,\,ij}^{V,LR}\right) \left(f_{+} P_{\mu}  + \left(f_{0} - f_{+}\right)\frac{m_P^2-m^2}{q^2} q_{\mu}  \right)
 \left[ \overline{u_N} \gamma^{\mu} P_R v_{N'} \right]\,, \\
 \mathcal{M} \left(P \to V N\overline{N}\right)  &=  \bigg \lbrace \left(c_{qN,\, ij}^{V,RR} + c_{qN,\,ij}^{V,LR}\right) \left( i g \epsilon_{\mu \nu \alpha \beta} \epsilon^{\ast \nu} p'^{\alpha} p^{\beta} \right)  \nonumber \\
&+ \frac{1}{2} \left(c_{qN,\, ij}^{V,RR} - c_{qN,\,ij}^{V,LR}\right) \left( f \epsilon^{\ast}_{\mu} + a_{+} P_{\mu} \epsilon^\ast \cdot p + a_- q_{\mu} \epsilon^\ast \cdot p  \right) \bigg \rbrace
 \left[ \overline{u_N} \gamma^{\mu} P_R v_{N'} \right]\,,
\end{align}
%%%%%%%%%%%%%%%%%%%%%%%%%
where $P'$ and $V$ are the pesudoscalar and vector mesons. In semileptonic meson decays, the hadronic matrix elements are described by the corresponding form factors.  If it is a pseudoscalar, the non-vanishing matrix element is:
%%%%%%%%%%%%%%%%%%%%%%%%%%%
\begin{align}
\langle P' (p') | \overline{q_i} \gamma^\mu q_j | P(p) \rangle &= f_{+}(q^2) \left[ (p + p')^{\mu} - \frac{m_P^2-m^2}{q^2} q^{\mu}\right] + f_{0}(q^2) \frac{m_P^2-m^2}{q^2} q^{\mu}, \,
\end{align} 
where $q^{\mu} \equiv p^{\mu} - p'^{\mu}$. When the final meson is a vector, the non-zero matrix elements are:
%%%%%%%%%%%%%%%%%%
\begin{align}
\langle V (p',\epsilon) | \overline{q_i} \gamma^{\mu} q_j | P(p) \rangle &= i g(q^2) \epsilon^{\mu \nu \alpha \beta} \epsilon_{\nu}^{\ast} P_{\alpha} q_{\beta} \,, \nonumber\\ 
\langle V (p',\epsilon) | \overline{q_i} \gamma^{\mu} \gamma_{5} q_j | P(p) \rangle &= f(q^2) \epsilon^{\ast \mu} + a_{+}(q^2) P^{\mu} \epsilon^{\ast} \cdot p + a_{-}(q^2) q^{\mu} \epsilon^{\ast} \cdot p \,, 
\end{align}
where we defined $P^{\mu} \equiv p^{\mu} + p'^{\mu}$. For these form factors we followed the Refs.~\cite{FlavourLatticeAveragingGroupFLAG:2024oxs,Lubicz:2017syv,Ivanov:2019nqd,Wu:2006rd,Bharucha:2015bzk}.
Then, the spin-summed 
matrix element squared can be written as
\begin{equation}
\sum_\mathrm{spins} \left|\mathcal{M}\right|^2 \bigg \rvert_{p_b=q-p_a} = \sum_{n=0}^K c_n(a) (p_a \cdot p)^n \,,
\label{eq:matrixelement}
\end{equation}
%%%%%%%%%%%%%%%%%%%%%%%%%
where $c_n(a)$ are functions of $a$, particle masses and the hadronic form factors. For the decays under study we have $K \leq 2$. After performing the remaining momentum integrals, one arrives at the final decay rate expression
%%%%%%%%%%%%%%%%%%%%%%%%%%%%%%%
where $p$ ($p'$) is the momentum of the decaying (outgoing) meson and $p_a$ and $p_b$ denote the momenta of the produced leptons.%
%%%%%%%%%%%%%%%%%%%%%%%%%%%%%%%%
\begin{equation}
\Gamma = \frac{1}{2m_P} \frac{1}{(2\pi)^5} \int^{(m_P-m)^2}_{(m_a+m_b)^2} \mathrm{d}a \, I_P \sum^K_{n=0} c_n(a) I_n \,,
\label{Gamma3body}
\end{equation}
where %the results of 
the integrals $I_P$ and $I_n$ (for $n={0,1,2}$) are given by
%%%%%%%%%%%%%%%%%%%%%%%
\begin{align}
I_P &= \frac{\pi}{2m_P^2} \lambda^{1/2}\left(a,m^2,m_P^2\right) \,, \,\,\,
I_0 = \frac{\pi}{2a} \lambda^{1/2}\left(a,m_a^2,m_b^2\right) \,, \,\,\,
I_1 = \frac{(p_a \cdot q)(p\cdot q)}{a} I_0 \,, \nonumber \\
I_2 &= \frac{I_0}{a^2} \left ( (p_a \cdot q)^2(p\cdot q)^2 + \frac{1}{48} \lambda(a,m^2,m_P^2) \lambda(a,m_a^2,m_b^2) \right) \,.
\end{align}
where $\lambda \left(x,y,z \right) \equiv x^2+y^2+z^2-2x y-2x z-2 y z$. The scalar products above have to be replaced by 
\begin{equation}
p_a \cdot q = \frac{1}{2} \left(a + m_a^2 - m_b^2\right) \,, \qquad p\cdot q = \frac{1}{2} \left(m_P^2 + a - m^2 \right) \,.
\end{equation} 
%%%%%%%%%%%%%%%%%%%%%%%%%%%%%
\bibliographystyle{utphys}
\bibliography{bibliography.bib}
\end{document}